\documentclass[prb,eqsecnum,noshowpacs]{revtex4}
\usepackage{amssymb,amsfonts,amsmath,bm,dcolumn,latexsym,graphicx}
\font\bba=msbm10 scaled 1200
\font\bbb=msbm8 %scaled 1080
\font\bbc=msbm6 %scaled 1080
\newfam\bbfam
\textfont\bbfam=\bba
\scriptfont\bbfam=\bbb
\scriptscriptfont\bbfam=\bbc
\def\bb{\fam\bbfam\bba}
\def\R{{\bb R}}

\begin{document}
\title{New mean field theories for the liquid-vapor transition 
\\ of charged hard spheres.}
\author{Jean-Michel Caillol}
\affiliation{Laboratoire de Physique Th\'eorique \\
UMR 8267, B\^at. 210 \\
Universit\'e de Paris-Sud \\
91405 Orsay Cedex, France}
\email{Jean-Michel.Caillol@th.u-psud.fr}
\date{\today}
\begin{abstract}
The phase behavior of the 
primitive model of electrolytes is
studied in the framework of various mean field
approximations obtained recently by means of methods pertaining to
statistical field theory
(CAILLOL, J.-M., 2004, 
\textit{J. Stat. Phys.}, \textbf{115}, 1461).
The role of the regularization of  the Coulomb potential at
short distances is discussed in details and
the link with more traditional approximations of the theory of  liquids
is discussed. 
The values computed for the critical temperatures, chemical
potentials, and densities are compared with  available Monte Carlo data and
other theoretical predictions. 
\end{abstract}
%\pacs{05.70.Fh,05.70.Jk,02.70.Rr,61.2.Qg,64.70.Fx}
\maketitle
%%%%%%%%%%%%%%%%%%%%%%%%%%%%%%%%%%%%%%%%
\section{\label{intro}Introduction}
Various ionic systems including electrolyte solutions, molten salts, and 
collo\"{i}ds can be studied with a good approximation
in the framework of the so-called primitive model (PM) which
consists in a neutral mixture of  $M$ species of charged hard spheres (HS)
of charges $q_{\alpha}$ and diameters $\sigma_{\alpha}$
($\alpha=1, \ldots, M $)\cite{Hansen}. Of special interest is the restricted
primitive model (RPM) where $M=2$, the spheres have all the same diameter 
$\sigma$, and the cations and anions bear opposite charges $\pm q$.
In this paper we shall also consider  the special primitive model (SPM)
in which the number $M$ of species as well as the charges $q_{\alpha}$
are arbitrary but all the ions have the same diameter $\sigma$. The case of a
size assymetry of the ions will not be considered however.

We have presented elsewhere an exact field theoretical representation of the PM
obtained by performing a Kac-Siegert-Stratonovich-Hubbard-Edwards 
(KSSHE) transformation
\cite{Kac,Siegert,Strato,Hubbard1,Hubbard2,Edwards,Brydges}
of the Boltzmann's factor \cite{Cai-Raim,Raim-Cai,Cai-JSP}. Thanks to this
transformation, the
grand-canonical partition function (GCPF) of the PM is essentially equal
to the GCPF of a
fluid of bare hard spheres in the presence of an imaginary  random Gaussian
field $i \varphi$ 
($ \varphi$ real) with a 
covariance given by the Coulomb potential. 
Since the  action  ${\cal H}[\varphi]$ of the KSSHE field theory 
can be explicitely obtained then
all the sophisticated techniques developed in
statistical field theory\cite{Ma,Zinn} 
can  be applied \textit{ a priori} to the PM. 
Note that a KSSHE formalism can also be introduced to deal with the case of
neutral fluids \cite{Cai-Mol, Efimov}; the method is thus quite general
and has even been employed in numerical simulations \cite{Parinello}.
Besides  KSSHE formalism, various  field-theoretic approaches 
have been proposed recently to study lattice and off-lattice
versions of the PM. It is the place to cite the works by Di Caprio
\textit{et al.} \cite{Badiali1,Badiali2},
the phenomenological (coarse-grained) field theory of Ciach and
Stell \cite{Ciach1,Ciach2,Ciach3}, and recent applications of the method of
collective variables by Patsahan and Mryglod
\cite{Patsahan0,Patsahan1,Patsahan2}.

In ref.~(\onlinecite{Cai-JSP}) we have succeeded  in obtaining
the free energy  of the homogeneous SPM 
at  second order loop order in the framework of  KSSHE
field theory. This expression  is used here to construct a
Landau theory \cite{Chaikin} for  the
liquid-vapor (LV) transition of the model.
Many mean-field (MF) like theories of the LV transition of the RPM
or the SPM have
been proposed and are reviewed for instance in ref.~(\onlinecite{Schroer}).
The aim of building  MF theories similar to those discussed in this paper
is not only 
to attempt to reproduce more or less accurately
the LV coexistence curve of the RPM
and its critical point, but also to provide 
a good starting point for a possible renormalization group analysis of ionic
criticality. An interesting step in this direction was made recently in 
ref.~(\onlinecite{Patsahan1}). 

Our paper is organized as follows. 
In sec.~\ref{Prole} we resume and extend the analysis of
ref.~(\onlinecite{Cai-JSP}) on the KSSHE representation of the SPM.  Then we
specialize to the case of a binary SPM in sec.~\ref{Binary} where we give the
expressions of the Landau function at the second order
in the loop-expansion. This
yields various MF theories for the LV transition which are  studied numerically
in sec.~\ref{LG}. We conclude in  sec.~\ref{Conclusion}.

%%%%%%%%%%%%%%%%%%%%%%%%%%%%%%%%%%%%%%%%
\section{\label{Prole}Prolegomena}
%%%%%%%%%%%%%%%%%%%%%%%%%%%%%%%%%%%%%%%%%%%%%%%%%%%%%%%%%%%%%%%%%%%%%%
%%%%%%%%%%%%%%%%%%%%%%%%%%%%%%%%%%%%%%%%%%%%%%%%%%%%%%%%%%%%%%%%%%%%%%
\subsection{\label{KSSHE}The KSSHE transform} 

We consider  the three dimensional (3D) version of the  
SPM. 
The particles live in a domain $\Omega \subset \R^3$ of volume 
$\vert \Omega \vert$.
We work in the grand canonical (GC) ensemble. Let $\mu_{\alpha}$ be
the chemical potential of the species $\alpha$. For convenience
we shall define the 
dimensionless chemical potential  $\nu_{\alpha}$ to be 
$\nu_{\alpha}=\beta \mu_{\alpha}$ ($\beta=1/k T$, $k$
Boltzmann's constant, $T$ temperature).

Denoting by $\bm{r}^{\alpha}_{i_{\alpha}} \in \Omega $ the position of the 
$i_{\alpha}$-th ion of species ${\alpha}$ we note that 
only  configurations $\omega \equiv 
(N_{1};\bm{r}_{1}^{1}, \ldots, \bm{r}^{1}_{N_{1}} \vert \ldots \vert
N_{\alpha };\bm{r}_{1}^{\alpha}, \ldots, \bm{r}^{\alpha}_{N_{\alpha}} \vert
\ldots \vert
N_{M};\bm{r}_{1}^{M}, \ldots, \bm{r}^{M}_{N_{M}} )$  
without overlaps of  the
spheres - i.e. such that  $\| \bm{r}^{\alpha}_{i_{\alpha}}
- \bm{r}^{\beta}_{i_{\beta}}\|  \geq \sigma$ -  
do contribute to the  GCPF $ \Xi[\beta, \{\nu_{\alpha}\},\vert \Omega \vert]$.
It follows from this remark that the ions can be supposed  to interact
via a pair potential $v_{\alpha,\beta}(r)=q_{\alpha}q_{\beta} \; w(r)$ where
$w(r)$ identifies of course with the Coulomb potential $1/r$ outside the HS
core, i.e. for $r \ge \sigma$, but is otherwise arbitrary inside 
(i.e. for $r \le \sigma$).

We have shown in previous works devoted either to neutral or charged fluids 
how to take advantage of the
arbitrariness of $w(r)$ inside the HS core to define properly the KSSHE
transform \cite{Cai-Raim,Raim-Cai,Cai-Mol,Cai-JSP}. In the case of the SPM
the crux of the whole matter is to rewrite the
electrostatic part of the configurational energy as a definite positive
quadratic form. Therefore the regularization of 
$w(r)$ for $0\leq r \leq \sigma$ must be chosen in such a way  
that the Fourier transform $\widetilde{w}(k)$ is
positive and $w(0)$ is finite.\cite{Brydges,Cai-Raim,Raim-Cai,Cai-JSP} 
Replacing the point charge $q_{\alpha}$ of each ion by  a 
radial distribution of charges  $q_{\alpha}\tau(r) $ smeared out inside 
its volume 
fulfills all these
requirements and proves to be convenient. Thus, assuming
\begin{eqnarray}
\label{tau}
\tau(r) & =&  0  \; \text{ if } r  \ge
 \overline{\sigma} \equiv \sigma/2
  \; , \nonumber \\
\widetilde{\tau}(k)& =&  1 \; ,
\end{eqnarray}
where $\widetilde{\tau}(k)$ denotes the Fourier transform of 
$\tau(r)$ we  indeed have
\begin{equation}
\label{wtilde}
\widetilde{w}(k)  = \frac{4 \pi \widetilde{\tau}^2(k)}{k^2}  \; \ge 0
\end{equation}
as required and, of course $w(r)=1/r$ for $r \ge \sigma$ by virtue of Gauss
theorem. Moreover for any reasonable distribution
 $\tau(r)$ the self-energy $q_{\alpha}^2 \; w(0)/2$ 
of ion $q_{\alpha}$ will be a finite quantity. 
For instance one can chose a uniform surface distribution of charges of radius
$\overline{a}=a/2 <\overline{\sigma}$, i.e.
 \begin{eqnarray}
\label{tau-a}
\tau_{a}(r)&=& \frac{1}{\pi a^2} \;
\delta(\| \bm{r}\| - \overline{a} ) \; , \nonumber \\
\widetilde{\tau_{a}}(k)&=& 
\frac{\sin k\overline{a}}{ k\overline{a}}
\; ,
\end{eqnarray}
yielding the regularized $w_{a}(r)$
\begin{eqnarray}
w_{a}(r) &=&   \frac{1}{r} \text{ for } r \geq a
 \; , \nonumber 
 \\
 \label{kk}
&=&   \frac{2 a -r }{a^2} \text{ for } r \leq a \; .
\end{eqnarray}
Other types of 
regularization will be discussed in the remainder of the paper.

Under these hypothesis a KSSHE transform can be performed and 
the GCPF of the SPM can be rewritten as \cite{Cai-JSP}
\begin{equation}
    \label{basic}
  \Xi[ \{\nu_{\alpha} \}]=  \left<
   \Xi_{\text{HS}}\left[ \{\overline{\nu}_{\alpha} +i q_{\alpha}\phi\}\right]
   \right>_{v_{c}}  \; ,  
\end{equation}  
where $\overline{\nu}_{\alpha} = \nu_{\alpha} +q_{\alpha}^2 \; w(0)/2$ is a
renormalized chemical potential and
$\Xi_{\text{HS}}\left[ \{\overline{\nu}_{\alpha} +i 
 q_{\alpha}\phi\}\right]$ is the GCPF of a mixture of bare
hard spheres in the presence of the local chemical potentials
 $\overline{\nu}_{\alpha} +i q_{\alpha}
\phi(\vec{r})$. The ''smeared'' field $\phi  $ is defined to be 
 the space convolution
\begin{equation}
    \label{Phi1}
 \phi(\bm{r}_{1})=  \beta^{1/2}  \int_{\Omega} d^{3} \bm{r}_{2} \;  
 \tau(\|\bm{r}_{1}-\bm{r}_{2}\|) \varphi(\bm{r}_{2}) 
 \end{equation}
 which will be  conveniently noted
 \begin{equation}
    \label{Phi2}
 \phi(1) =  \beta^{1/2} \;  
 \tau(1,2) \varphi(2) \; .
 \end{equation}
Note that in  the whole  paper,
summation over repeated  indices, either discrete or continuous, 
will always be meant (except if explicitly stated otherwise). 
 
The brackets $\left< \ldots \right>_{v_{c}}$ in equation ~(\ref{basic}) denote 
a Gaussian 
average over the real scalar field $\varphi(\bm{r})$, i.e.
\begin{eqnarray}
\label{moyv}
\left< \ldots  \right>_{ v_{c}}&\equiv&
{\cal N}_{v_{c}}^{-1} \int {\cal D} \varphi
\ldots \exp\left(-\frac{1}{2} \left< \varphi \vert  v_{c}^{-1}
\vert \varphi \right>\right) \; , \nonumber \\
{\cal N}_{v_{c}} &\equiv& \int {\cal D } \varphi
\exp\left(-\frac{1}{2} \left< \varphi \vert  v_{c}^{-1}
\vert \varphi \right>\right) \; ,
\end{eqnarray}
where ${\cal D}\varphi $ is the measure of the functional integration and
\begin{equation}
    \label{vmoins1}
 v_{c}^{-1}(1,2) = - \frac{1}{4 \pi} \Delta_{1}\delta (1,2) 
 \end{equation} 
is the inverse of the positive operator $v_{c}(1,2)\equiv 1/r_{12}$.
We have also made use of the convenient notation
\begin{eqnarray}
\label{nota}
 \left< \varphi \vert  v_{c}^{-1}
\vert \varphi \right>&= &\int_{\Omega} d^{3} \bm{r}_{1} 
 \int_{\Omega} d^{3} \bm{r}_{2} \; 
 \varphi(\bm{r}_{1})\;
  v_{c}^{-1}(\bm{r}_{1},\bm{r}_{2})\;
 \varphi(\bm{r}_{2})  \nonumber \\
 &\equiv & \varphi(1)v_{c}^{-1}(1,2) \varphi(2) \; .
\end{eqnarray}
After an integration by parts with appropriate boundary conditions 
one obtains the more transparent expression\cite{Brydges,Cai-JSP}
\begin{equation}
    \label{Nv}
 \left< \varphi \vert  v_{c}^{-1}
\vert \varphi \right> =      
     \frac{1}{4 \pi} \int_{\Omega} d^{3} \bm{r}\; ( \bm{\nabla} 
    \varphi )^{2}
   \; .
\end{equation}
To make  some contact with statistical field theory we finally  introduce the
KSSHE action or effective Hamiltonian
\begin{equation}
\label{H}
{\cal H}[\varphi]  = \frac{1}{2} \left< 
\varphi \vert  v_{c}^{-1} \vert \varphi \right>  - 
\ln \Xi_{\text{HS}}\left[ \{\overline{\nu}_{\alpha} +i q_{\alpha}\phi \}
\right] \; ,
\end{equation}
which allows us to recast $\Xi$ under the form
\begin{equation}
\label{Xinewlook}
\Xi[ \{\nu_{\alpha} \}]= {\cal N}_{v_{c}}^{-1} \int {\cal D} \varphi
\; \exp(-{\cal H}[\varphi])
\: .
\end{equation}
It is important 
to distinguish carefully, besides the usual GC averages  $<\ldots>_{\text{GC}}$,
between two types of statistical field
averages :
the already defined  $<\ldots>_{ v_{c}}$ (cf equation~(\ref{moyv}))
and the  $<\ldots>_{{\cal H}}$  defined as
\begin{equation}
<A[\varphi]>_{{\cal H}} \equiv \frac{\int {\cal D} \varphi \;
 \exp(-{\cal H}[\varphi])A[\varphi]}{\int {\cal D} \varphi
 \; \exp(-{\cal H}[\varphi])} \; .
\end{equation}
Since many thermodynamic quantities of interest can be expressed
in terms of the  charge correlation functions $G^{(n)}_\text{C}$ 
\cite{Hansen} it is
important to relate these  functions to the field
correlation functions $G^{(n)}_{\varphi}$. More precisely one has
\begin{subequations}
\label{G}
\begin{eqnarray}
\label{Gc}
G^{(n)}_\text{C}(1,\ldots,n)&=&<\widehat{\rho}_\text{C}(1)\ldots
 \widehat{\rho}_\text{C}(n)>_{\text{GC}}
 \; , \\
\label{Gphi} 
G^{(n)}_{\varphi}(1,\ldots,n)&=&<\varphi(1)\ldots \varphi(n) >_{{\cal H}}\; ,
\end{eqnarray}
\end{subequations}
where, in equation~(\ref{Gc}), the microscopic charge density 
$\widehat{\rho}_\text{C}$ is given by
\begin{eqnarray}
\label{roc}
\widehat{\rho}_\text{C}(1)&=&\tau(1,2)\; q_{\alpha}
\widehat{\rho}_{\alpha}(2)  \; , \nonumber \\
\widehat{\rho}_{\alpha}(\bm{r})&=&\sum_{i=1}^{N_{\alpha}} 
\delta^{(3)}(\bm{r}-
\bm{r}^{\alpha}_i) \; ,
 \end{eqnarray}
the summation over the dummy indices $\alpha$ being understood in the first line
of equation~(\ref{roc}). 
It proves convenient to define also the truncated (or connected)
 n-body correlation
functions by the relations
\begin{eqnarray}
\label{defcorreCT3}
G^{(1)\; \text{T}}_{\text{C} \;(\varphi)} (1) &=& G^{(1)}_{\text{C} \;(\varphi)} (1)
\; , \nonumber \\
G^{(n)\; \text{T}}_{\text{C} \;(\varphi)} (1, \ldots, n) &=&
G^{(n)}_{\text{C} \; (\varphi)} ( 1,\ldots,n)
- \sum \prod_{m<n}G^{(m)\; \text{T}}_{\text{C} \;(\varphi)} (i_{1},\ldots,i_{m}) \; 
\text{for } n\geq 2
 \; ,
\end{eqnarray} 
where the sum of products is carried out over all possible partitions of 
the set $(1,\ldots,n)$ into subsets of cardinality $m<n$ \cite{Percus}.
The relations between the truncated 
$G^{(n)\; \text{T}}_\text{C}$ and $G^{(n)\; \text{T}}_{\varphi}$ were obtained  in
ref.~(\onlinecite{Cai-JSP}) by a lengthy and cryptic method, a new simple
derivation is given in appendix~A. For 
a homogeneous system one finds at order $n=0$ the charge 
neutrality condition, i.e. 
 \begin{equation}
 \label{neutrality}
 \rho_{\text{C}} = \rho_{\alpha} q_{\alpha}=0 \; ,
 \end{equation} 
and, at higher orders, the following relations
\begin{subequations}
\label{GnGphi}
\begin{eqnarray}
 \beta G^{(2)\; \text{T}}_{\text{C}}(1, 2)&= &\frac{-1}{4 \pi} \Delta_{1} \delta(1,2)
 -  \frac{1}{(4 \pi)^{2}} \Delta_{1} \Delta_{2}
  G^{(2) \; \text{T}}_{\varphi}(1, 
 2) \; , \\
\beta^{n/2} i^n G^{(n)\; \text{T}}_\text{C}(1,\ldots,n)&=& 
\frac{(-1)^{n}}{(4 \pi)^{n}} \Delta_{1} \ldots  \Delta_{n}
  G^{(n) \; \text{T}}_{\varphi}(1, \ldots,n)  \; \; \text{ for } n \geq 3 \; . 
\end{eqnarray}
\end{subequations}

In ref.~(\onlinecite{Cai-JSP}) we have computed the free energy
$f(\{\rho_{\alpha}\})$ of the homogeneous SPM at the second loop order
in the framework of the KSSHE field theory. Before giving and 
discussing this expression
we want to stress that the exact $f(\{\rho_{\alpha}\})$ should be independent
of the form of the pair potential $w(r)$ inside the core, i.e. of the type 
of regularization adopted for the KSSHE transform. More generally,
each term of say a  systematic low fugacity or high temperature
expansion of $f(\{\rho_{\alpha}\})$
should also be independent of this regularization; 
this point was carefully checked
in refs.~(\onlinecite{Cai-Raim,Raim-Cai}) in the case of the RPM.
However here, as in ref.~(\onlinecite{Cai-JSP}),  we
consider a loop-wise expansion of $f(\{\rho_{\alpha}\})$. The small parameter 
involved in this expansion cannot be given a clear physical interpretation
and serves only to keep track of certain classes of 
Feynman diagrams \cite{Zinn}.
Consequently each term of the loop expansion depends upon  $w(r)$ inside
the core.
Deciding which type of regularization to adopt  is 
a matter of arbitrariness or mathematical skill. However it seems reasonable 
to impose that, at each order of
the loop expansion, $f(\{\rho_{\alpha}\})$ should be stationary with respect 
to the variations of  $w(r)$ inside the core.

The zero-loop (mean-field (MF) or tree level),
one-loop and two-loops expressions of
$f(\{\rho_{\alpha}\})$ will be denoted  by $f^{(0)}$,
$f^{(1)}$, and $f^{(2)}$ respectively. Although the expressions obtained for
$f^{(0)}$, $f^{(1)}$, and $f^{(2)}$ in ref.~(\onlinecite{Cai-JSP})
were derived with the assumption of a charge smearing regularization for $w(r)$
one can safely
assume that they remain
valid for any reasonable regularization of $w(r)$ inside the HS core
(i.e. such that 
$\widetilde{w}(k)>0$ and $w(r)=1/r$ for
$r>\sigma$). 
%%%%%%%%%%%%%%%%%%%%%%%%%%%%%%%%%%%%%%%%%%%%%%%%%%%%%%%%%%%%%%%%%%%
%%%%%%%%%%%%%%%%%%%%%%%%%%%%%%%%%%%%%%%%%%%%%%%%%%%%%%%%%%%%%%%%%%%
\subsection{\label{0l}Zero-loop order}
For a homogeneous
system one finds for the zero-loop free energy per unit volume
the sloppy result \cite{Cai-JSP}

\begin{equation}
\label{f-0loop}
\beta f^{(0)}(\{\rho_{\alpha}\})=
 \beta f_{\text{HS}}(\{\rho_{\alpha}\}) -\frac{\beta}{2}
 \; \rho_{\alpha}q_{\alpha}^{2} \; w(0)\; ,
\end{equation}
where  
$f_{\text{HS}}(\{\rho_{\alpha}\})$ denotes the excess free energy per unit
volume of 
the reference HS fluid. Moreover the charge neutrality condition 
$\rho_{\alpha}q_{\alpha}=0$ has to be be satisfied \cite{Cai-JSP}.
Note that $\beta f^{(0)}$ diverges to $- \infty$ for a point
like distribution $\tau(r)=\delta^{(3)}(\bm{r})$ but remains finite after
regularization if $w(r)$ is well behaved a $r=0$. 
Although disappointingly simple, 
equation~(\ref{f-0loop}) can be however exploited for it can be shown that 
the MF
free energy $\beta f^{(0)}(\{\rho_{\alpha}\})$ constitutes an exact lower
bound for the exact free energy $f(\{\rho_{\alpha}\})$ \cite{Cai-JSP}.
Maximizing the expression~(\ref{f-0loop}) of $\beta f^{(0)}(\{\rho_{\alpha}\})$
with respect to he variations of $w(r)$ inside the core should yield  an
optimized lower bound for  $f(\{\rho_{\alpha}\})$; unfortunately this
mathematical problem has no solution  since,
 as apparent from equation~(\ref{f-0loop}), 
$\beta f^{(0)}(\{\rho_{\alpha}\})$ is a linear functional of $w(r)$.
However, if
one restricts oneself to potentials of the form~(\ref{wtilde}) 
the problem can be
solved \cite{Cai-JSP} and one finds that the distribution $\tau(r)$ 
which maximizes
$\beta f^{(0)}(\{\rho_{\alpha}\})$ is a uniform surface distribution of charges
of radius
equal to that of the ions, i.e. the distribution $\tau_{\sigma}(r)$ 
given by  equation~(\ref{tau-a}) with $a \equiv \sigma$ ,  
whence the optimized lower bound 
$ \beta f_{\text{HS}}(\{\rho_{\alpha}\}) -\beta
\rho_{\alpha}q_{\alpha}^{2}/\sigma $ for the free energy, i.e.
nothing but the well known Onsager bound \cite{Hansen,Onsager}.
We christen this
regularization scheme of $w(r)$ as the optimized mean field (OMF) scheme.

The pair correlation functions $h_{\alpha,\beta}^{(0)}(r)\equiv
g_{\alpha,\beta}^{(0)}(r)-1 $ and the direct 
correlation
functions  $c_{\alpha,\beta}^{(0)}(r)$ \cite{Hansen}  at the  zero-loop
order are related to the free propagator $\Delta(r)$ of  KSSHE
field theory. In Fourier space one has \cite{Cai-JSP}

\begin{subequations}
\label{RPA}
\begin{eqnarray}
\widetilde{h}_{\alpha,\beta}^{(0)}(k)&=& \widetilde{h}_{\text{HS}, \; \rho}(k) 
- \beta 
q_{\alpha} q_{\beta} \widetilde{\Delta}(k) \; , \\
\widetilde{c}_{\alpha,\beta}^{(0)}(k)&=& \widetilde{c}_{\text{HS}, \;  \rho}(k)
 - \beta 
q_{\alpha} q_{\beta} \widetilde{w}(k) \; ,
\end{eqnarray} 
\end{subequations}
where $\widetilde{h}_{\text{HS}, \; \rho}(k)$ and 
 $\widetilde{c}_{\text{HS}, \;\rho}(k)$ denote the
Fourier transforms of the ordinary and direct correlation functions
of a HS fluid
at the density $\rho=\sum_{\alpha} \rho_{\alpha}$ respectively. We emphasize
that   $h_{\alpha,\beta}^{(0)}(r)$ and $c_{\alpha,\beta}^{(0)}(r)$, as given
by equations~(\ref{RPA}), do satisfy to the Ornstein-Zernicke (OZ) relations
 \cite{Hansen}
\begin{equation}
\label{OZ} 
h_{\alpha,\beta}^{(0)}(r)=c_{\alpha,\beta}^{(0)}(r) +
\rho_{\gamma} c_{\alpha,\gamma}^{(0)}\star h_{\gamma,\beta}^{(0)}(r) \;
\end{equation}
where the symbol "$\star$" denotes a convolution in space.
The  Fourier transform of the propagator
has the following expression
 \cite{Cai-JSP}
 \begin{equation}
 \label{Deltak}
 \widetilde{\Delta}(k)=\frac{\widetilde{w}(k)}{1+\beta \;
 \rho_{\alpha}q_{\alpha}^{2} \;\widetilde{w}(k) }.
\end{equation}
Therefore $w(r)$ and $\Delta(r)$ also satisfy to an OZ-like equation in direct
space, that is 
\begin{equation}
\label{Deltar}
\Delta(r) = w(r) - \beta  \;[\rho_{\alpha}q_{\alpha}^{2}] \;
 \Delta \star w (r) \; .
\end{equation}
We stress that the propagator $\Delta(r)$ as well as  the correlation functions
 $h_{\alpha,\beta}^{(0)}(r)$ and $c_{\alpha,\beta}^{(0)}(r)$ do
 depend on the form of $w (r)$ inside the core.
%%%%%%%%%%%%%%%%%%%%%%%%%%%%%%%%%%%%%%%%%%%%%%%%%%%%%%%%%%%%%%%%%%%%%%
%%%%%%%%%%%%%%%%%%%%%%%%%%%%%%%%%%%%%%%%%%%%%%%%%%%%%%%%%%%%%%%%%%%%%%
\subsection{\label{1l}One-loop order} 
 The one-loop free energy per unit volume is given by \cite{Cai-JSP}
 \begin{eqnarray}
\label{loop-1}
\beta f^{(1)}(\{\rho_{\alpha}\})&=&\beta f^{(0)}(\{\rho_{\alpha}\}) 
+ \frac{1}{2}\int \frac{d^{3}\bm{k}}{(2 \pi)^{3}}\; 
\ln(1+ \beta [\rho_{\alpha}q_{\alpha}^{2}] \widetilde{w}(k)) 
 \nonumber \\
 &=& \beta f_{\text{HS}}(\{\rho_{\alpha}\}) + \frac{1}{2}
 \int \frac{d^{3}\bm{k}}{(2 \pi)^{3}}\;
 \left[  \ln\left(1+ \beta [\rho_{\alpha}q_{\alpha}^{2}] \widetilde{w}
 \left(k\right)\right)-
 \beta  [\rho_{\alpha}q_{\alpha}^{2}] \widetilde{w}\left(k\right) \right] 
 \; . 
\end{eqnarray}
 $\beta f^{(1)}$ obviously still depends on the pair potential $w(r)$
 inside the core but  we note by passing
 that  the expression~(\ref{loop-1}) of $ \beta f^{(1)}(\{\rho_{\alpha}\})$
 remains
 finite (contrary to that of $\beta f^{(0)}$) for point-like  distributions
 of charges, i.e. for $\tau_{\text{DH}}(r)=\delta^{(3)}(\bm{r})$.
 One indeed obtains in this case the well-known Debye-H\"uckel (DH) result
 \begin{equation}
 \label{DH}
 \beta f_{\text{DH}}(\{\rho_{\alpha}\})=\beta f_{\text{DH}}(\{\rho_{\alpha}\})
 - \frac{\kappa^{3}}{12 \pi} \; ,
\end{equation}
where $\kappa^{2}=4\pi\beta \; \rho_{\alpha}q_{\alpha}^{2} $ 
is the squared Debye number. Incidently the DH propagator can also be computed
with the result
\begin{equation}
 \label{Delta-DH} 
 \Delta_{\text{DH}}(r)=\exp(-\kappa r)/r \; .
\end{equation}

In the Gaussian approximation  
the free energy is given
by its one-loop expression~(\ref{loop-1}) and the correlation functions by their 
zero-loop expressions~(\ref{RPA}) \cite{Cai-JSP,Ma}.
The Gaussian approximation of the KSSHE field theory therefore
coincides exactly with the usual random phase approximation (RPA) of the 
theory of liquids \cite{Hansen,Andersen}. One can further demand  the
one-loop free energy $\beta f^{(1)}(\{\rho_{\alpha}\})$ to be independent of
the expression of $w(r)$ inside the core. As well known the stationary condition
\begin{equation}
\frac{\delta \beta f^{(1)}(\{\rho_{\alpha}\})}{\delta w(\bm{r})}=0 \text{ for }
r\leq \sigma \; ,
\end{equation}
yields the conditions $g_{\alpha,\beta}^{(0)}(r)$ for $0 \leq r\leq \sigma$
and for all  pairs $(\alpha,\beta)$. We recover in that way 
the optimized random phase
approximation (ORPA) of the theory of liquids \cite{Andersen,Hansen}.
It follows
from eqs.~(\ref{RPA}) that the condition of nullity of 
 $g_{\alpha,\beta}^{(0)}(r)$ inside the core can be rewritten as 
 \begin{eqnarray}
 \label{MSA-cond}
 w(r)&=& 1/r \text{ for } r\geq \sigma \nonumber \\
 \Delta (r)& =& 0 \text{ for } r\leq \sigma \; ,
 \end{eqnarray}
where $w(r)$ and $\Delta (r)$ are linked by the integral
equation~(\ref{Deltar}). Note that this equation does not involve the
correlation functions of the reference HS system and that
it also appears in the mean
spherical approximation (MSA) of the SPM (in this case the exact correlation
functions $h_{\text{HS}}$ and  $c_{\text{HS}}$ 
in the right hand side  of eqs.~(\ref{RPA})
have to be replaced by their Percus-Yevick (PY)  expressions
, i.e.
 $h_{\text{PY}}$ and  $c_{\text{PY}}$ respectively
 \cite{Hansen}).
It turns out that equations~(\ref{Deltar}) and~(\ref{MSA-cond})  
can be solved analytically with the result\cite{Lebo,Hansen}
\begin{subequations}
 \begin{eqnarray}
 \label{MSA}
  w_{\text{MSA}}(r)&=& 1/r    \text{ for } r\geq \sigma \; , \\
  \label{jj}
  w_{\text{MSA}}(r)&=& \frac{B}{\sigma} \; (2- \frac{B r }{\sigma})   
  \text{ for } 0 \leq r \leq \sigma  \; ,               \\
  B&=& \frac{x^2 +x -x(1+2 x)^{1/2}}{x^2} \; ,
\end{eqnarray}
\end{subequations}
where $x=\kappa \sigma$. We remark that the function $w_{\text{MSA}}(r)$ 
meets all the requirements
of a regularized potential, i.e. $w_{\text{MSA}}(r)=1/r$ for $r\geq \sigma$ and
$\widetilde{w}(k) \geq 0$ as  can be shown easily, and therefore
can be used as a regularizator of the KSSHE transform. It must be
stressed that, despite of
the formal analogies between equations~(\ref{jj}) and~(\ref{kk}) -the latter
being deduced from the former by setting $B=1$-, 
 $w_{\text{MSA}}(r)$
cannot be interpreted as the interaction energy between 
two smeared distributions of
charge since $0<B<1$ as discussed in ref.(\onlinecite{Cai-JSP}).
The same function  $w_{\text{MSA}}(r)$ 
appears in the MSA
and ORPA approximations for the SPM and yields in both cases ion-ion
 pair correlation
functions which vanish inside the HS core.  By contrast, the regularized potential
$w_{\sigma}(r)$ which enters the OMF free energy
does not insure the nullity of $g_{\alpha,\beta}^{(0)}(r)$ for 
$r \leq \sigma$. 

With the choice $w(r)\equiv w_{\text{MSA}}(r)$ the one-loop free
energy~(\ref{loop-1})
can be computed exactly. The result is well-known \cite{Lebo} but we cannot
refrain from giving the following
derivation which seems to be original. Let us first
compute the internal energy per unit volume
 $u_{\text{ORPA}}=\partial \beta f_{\text{ORPA}}(\{\rho_{\alpha}\}) 
 /\partial \beta$. It
follows from equation~(\ref{loop-1}) that
\begin{eqnarray}
\label{u-ORPA} 
u_{\text{ORPA}} &=& -\frac{1}{2} \; 
\int \frac{d^{3}\bm{k}}{(2 \pi)^{3}}\;
\beta \; \left[\rho_{\alpha}q_{\alpha}^{2}\right] \;  
\widetilde{\Delta}_{\text{MSA}}(k) \;
\frac{\partial}{\partial\beta }
\left[
\beta \; \left[\rho_{\alpha}q_{\alpha}^{2}\right] \; 
 \widetilde{w}_{\text{MSA}}(k) \right]\nonumber \\
  &=& -\frac{1}{2}
\int d^{3}\bm{r}\;\beta \; [\rho_{\alpha}q_{\alpha}^{2}] \;
\Delta_{\text{MSA}}(r)  
\frac{\partial}{\partial\beta }\left[\beta \; \left[
\rho_{\alpha}q_{\alpha}^{2}\right]
\;w_{\text{MSA}}(r)\right] \;
,  
\end{eqnarray}
where we made use of the expression~(\ref{Deltak}) of $\widetilde{\Delta}(k)$.
 Since the
propagator $\Delta_{\text{MSA}}(r)$ vanishes inside the core in the ORPA
(and MSA) approximations
we have merely to compute the derivative of
 $\beta \;[\rho_{\alpha}q_{\alpha}^{2}]\;w_{\text{MSA}}(r)$ 
 with respect to $\beta$ for
$r>\sigma$ which is trivial since $w_{\text{MSA}}(r)=1/r$ in this case.  
Therefore 
\begin{eqnarray}
\label{u-ORPA2}
u_{\text{ORPA}} &=& -\frac{\beta}{2} \; [\rho_{\alpha}q_{\alpha}^{2}]  \;
\int d^{3}\bm{r}\;\Delta_{\text{MSA}}(r) \;  [\rho_{\alpha}q_{\alpha}^{2}] 
\; w_{\text{MSA}}(r)\nonumber \\ 
 &=& \frac{\beta}{2} \; [\rho_{\alpha}q_{\alpha}^{2}] \left(\Delta_{\text{MSA}}
 \left(r=0 \right)-
w_{\text{MSA}}\left(r=0\right) \right) \; ,
\end{eqnarray}
where we made use of equation~(\ref{Deltar}) to obtain the last line.
Since $\Delta_{\text{MSA}}(0)=0$, we have finally 
 \begin{eqnarray}
 \label{u-ORPA3}
u_{\text{ORPA}}& = &\frac{-\beta}{2}\;  [\rho_{\alpha}q_{\alpha}^{2}]\;
w_{\text{MSA}}(0) \; , \nonumber \\
&=&  -\beta \; [\rho_{\alpha}q_{\alpha}^{2}]\; B / \sigma   \; ,
 \nonumber \\
&=& -\frac{x^2 + x -x(1+2 x)^{1/2}}{4 \pi\beta \sigma^3 }
\; ,
\end{eqnarray}
which  leads, after integration,  
to the ORPA free energy\cite{Hansen,Andersen,Lebo}
\begin{eqnarray}
\label{f-ORPA}
\beta f_{\text{ORPA}}(\{\rho_{\alpha}\}) &=&\beta f_{\text{HS}}(\{\rho_{\alpha}\}) 
+ \int_0^{\beta} d \beta^{'} \; u_{\text{ORPA}}(\beta^{'}) \nonumber \\
 &=&\beta f_{\text{HS}}(\{\rho_{\alpha}\}) -
 \frac{3 x^2 + 6 x +2 -2 (1+ 2 x)^{1/2}}{12 \pi \sigma^3} \; .
\end{eqnarray}

%%%%%%%%%%%%%%%%%%%%%%%%%%%%%%%%%%%%%%%%%%%%%%%%%%%%%%%%%%%%%%%%%%%%%%
%%%%%%%%%%%%%%%%%%%%%%%%%%%%%%%%%%%%%%%%%%%%%%%%%%%%%%%%%%%%%%%%%%%%%%
\subsection{\label{2l}Two-loop order} 
The free energy of the SPM at the second-loop order has the following
expression \cite{Cai-JSP}
\begin{eqnarray} 
\label{f-2loop} 
 \beta f^{(2)}(\{\rho_{\alpha}\})&=&
 \beta f^{(1)}(\{\rho_{\alpha}\}) - \frac{\beta^{2}}{4}
 [\rho_{\alpha}q_{\alpha}^{2}]^{2} \int  d^{3}\bm{r}\; 
 h_{\text{HS},\;\rho }(r) \Delta^{2}(r)    \nonumber \\
&+&\frac{\beta^{3}}{12} [\rho_{\alpha}q_{\alpha}^{3}]^{2} 
\int  d^{3}\bm{r} \;  \Delta^{3}(r) \;   
 \; ,
\end{eqnarray} 
where $ h_{\text{HS},\;\rho }(r)$ denotes the usual pair correlation function
of a
fluid of a single species of hard spheres at the total number density 
$\rho = \sum_{\alpha} \rho_{\alpha}$.
Demanding the independence of $\beta f^{(2)}(\{\rho_{\alpha}\})$ with respect
to the variations of $w(r)$ inside the core is a formidable mathematical task
that cannot be achieved analytically. Each of the regularization schemes that we have
discussed previously can be considered however, yielding different expressions
for the propagator $\Delta(r)$ and thus for 
$\beta f^{(2)}$. Expression~(\ref{f-2loop}) remains however
tractable, at least numerically, since it involves integrals of the functions
$\Delta(r)$ and $h_{\text{HS},\;\rho }(r)$ which can be both evaluated 
numerically 
(for instance, the approximation
 $h_{\text{HS},\;\rho}(r) \sim h_{\text{PY},\;\rho}(r)$
can be used safely at low and  moderate densities).  

%%%%%%%%%%%%%%%%%%%%%%%%%%%%%%%%%%%%%%%%%%%%%%%%%%%%%%%%%%%%%%%%%%%%%%
%%%%%%%%%%%%%%%%%%%%%%%%%%%%%%%%%%%%%%%%%%%%%%%%%%%%%%%%%%%%%%%%%%%%%%
\subsection{\label{WCA}The WCA regularization scheme}  
We end this section by discussing briefly the popular
Weeks-Chandler-Andersen (WCA) regularization scheme.
In the WCA scheme the potential $w(r)$ inside the core is assumed to be 
a constant equal to  $1/\sigma$ \cite{Weeks}. Clearly it amounts
to define $w(r)$ as the electric potential created by
a uniform surface distribution of charge $\tau_{\sigma}(r)$
where $\tau_{\sigma}(r)$ is the distribution defined at
equation~(\ref{tau-a}) with $a=\sigma$. 
It follows from this remark that the Fourier transform of the regularized pair
potential reads as
$$
\widetilde{w}_{\text{WCA}}(k)  = \frac{4 \pi \widetilde{\tau}_{\sigma}(k)}{k^2}
$$
with
$\widetilde{\tau}_{\sigma}=\sin(k \overline{\sigma})/(k\overline{\sigma})$.
Therefore $\widetilde{w}_{\text{WCA}}(k)$ is not positive for all $k$ and the
KSSHE transformation is ill-defined. As a consequence the
propagator $\Delta_{\text{WCA}}(r)$ can be singular (cf equation~(\ref{Deltak})).
This singularity yields the so-called RPA catastrophe\cite{Hansen} for the
one-loop (RPA)  free energy since the argument
of the $\ln$ in the RHS of equation~(\ref{loop-1}) can become negative. 
This finite wave-number $k$ instability,
which however is absent in the OMF and MSA
regularization schemes, has been  made responsible of the  possible  
existence of
charge-ordered phase \cite{Ciach1,Ciach2,Ciach3,Patsahan2}.
 Although an order-disorder transition of the
lattice-version of the RPM was indeed observed in Monte Carlo simulations
\cite{Pana2,Pana3}, the evidences
of  a similar transition for the off-lattice version of the model are 
lacking at the time of writing as far as the author is well informed.
   
%%%%%%%%%%%%%%%%%%%%%%%%%%%%%%%%%%%%%%%%%%%%%%%%%%%%%%%%
%%%%%%%%%%%%%%%%%%%%%%%%%%%%%%%%%%%%%%%%%%%%%%%%%%%%%%%%
\section{\label{Binary}The binary mixture}
In this section we specialize the results of section~\ref{Prole} to 
the case of a binary
mixture ($M=2$). The cations and the anions have all
the same diameter $\sigma$ and
bear charges $q_1=q$ and $q_2=-\xi q$ respectively where $\xi$ is the charge
assymetry parameter.
 It follows from the neutrality
condition~(\ref{neutrality}) that the ion densities are
$\rho_1=\xi\rho/(1+\xi)$ and $\rho_2=\rho/(1+\xi)$ where $\rho=\rho_1+\rho_2$
is the total number density.
 For convenience we introduce the reduced densities
 $\rho^{\star}_{\alpha}=\rho_{\alpha} \sigma^3$ and free energies per unit
 volume $\overline{\beta f}=\beta f \sigma^3$. The reduced 
 free energy $\overline{\beta f}_{\text{HS}}$
 of the reference HS system is thus given by
\begin{equation}
 \label{fHS}
\overline{\beta f}_{\text{HS}}(\rho_1^{\star},\rho_2^{\star})=
\overline{\beta f}_{\text{HS}}(\rho^{\star}) + \rho^{\star} \left( 
\frac{\xi}{1+\xi} \ln\xi  -\ln \left(1+\xi\right) 
  \right)\; ,
\end{equation}
where $\overline{\beta f}^{\text{HS}}(\rho^{\star})$ is the free energy 
of a pure HS liquid  at the  reduced density $\rho^{\star}$.
The additional term in the RHS
of equation~(\ref{fHS}) corresponds to the mixing entropy contribution.
 Following
the authors of ref.~(\onlinecite{Pana-Fisher}) we define the reduced inverse
temperature $\beta^{\star}\equiv \beta q^2 \xi/\sigma $ and the
reduced temperature as  $T^{\star}=1/\beta^{\star}$.
The reduced Debye number will of course be defined as
$\kappa^{\star} =\kappa \sigma$ (note that $\kappa^{\star 2}=4 \pi 
\beta^{\star}\rho^{\star}$) and  we will also frequently make 
use of the  dimensionless squared Debye number $\lambda=\kappa^{\star 2}  $.
 
Quite remarkably the electrostatic contribution to the one-loop free energy
depends solely on  parameter $\lambda$ and \textit{not} on the charge
 asymmetry factor  
$\xi$. Indeed one has 
\begin{equation}
\label{f1red}
\overline{\beta f}^{(1)}=\overline{\beta f}_{\text{HS}}(\rho^{\star})
 + \rho^{\star} \left( 
\frac{\xi}{1+\xi} \ln \xi  -\ln \left(1+\xi\right) \right)+
I[\lambda] \; ,
\end{equation}
where the expression of the function $I[\lambda]$ depends on the choice of
regularization made for $w(r)$.
The expressions of $I[\lambda]$ in the DH and ORPA
schemes have been given at eqs.~(\ref{DH}) and~(\ref{f-ORPA}) respectively. In
the OMF and WCA schemes one has 
\begin{subequations}
\label{I0}
\begin{eqnarray}
\label{I0-OMF}
I_{\text{OMF}}[\lambda]&=&  
\frac{2}{\pi^2} \int_0^{\infty}dx 
 \; x^2 \; \left(\ln\left( 1 + \frac{\lambda}{4} \frac{\sin x^4}{x^4}
 \right) -  \frac{\lambda}{4} \frac{\sin x^4}{x^4}\right) \;,
    \\
\label{I0-WCA}
     I_{\text{WCA}}[\lambda]             &=&\frac{1}{4 \pi^2}
 \int_0^{\infty}dx \; x^2
 \; \left(\ln\left( 1 + \frac{\lambda}{x^3}\sin x 
 \right) -  \frac{\lambda}{x^3}\sin x \right)		  
		        \;,
\end{eqnarray}
\end{subequations}
respectively. 
Both integrals occurring in eqs~(\ref{I0}) are convergent at $x=0$ and
$x=\infty$ and can be handled numerically easily.  However, the second
integral~(\ref{I0-WCA}) is singular for large values of $\lambda$ 
(the so-called RPA catastrophe)
since the argument
of the $\ln$ can become negative at finite $x$. The stability limit
is found to be  $\lambda \leq \lambda_{\text{max.}} \sim 84.2$.
 
In the case of a regularization
by a surface distribution of charges of radius $\overline{a}\equiv a/2
 \leq \overline{\sigma}$ one
has simply $I_{a}[\lambda]=I_{\sigma}[\lambda \overline{a}
^2]\overline{a}^{-3}
\equiv I_{\text{OMF}}[\lambda \overline{a}^2]\overline{a}^{-3}$. It is
easy to check that in the point-like limit $ a \to 0$ one gets
$I_{a}[\lambda]\to -
\lambda^{3/2}/12$, which coincides with the DH result~(\ref{DH}) as expected.

 At the two-loop order the free energy depends now not only on $\lambda$ but
 also on the density and the charge asymmetry parameter $\xi$. One deduces from
 equation~(\ref{f-2loop}) that
 
\begin{equation}
\label{f2red}
\overline{\beta f}^{(2)}= \overline{\beta f}^{(1)}
 -\frac{1}{4} K[\lambda,\rho^{\star}]
 + \frac{1}{12} \frac{(1-\xi)^2}{\xi} L[\lambda,\rho^{\star}]
\; ,
\end{equation}
with
 \begin{subequations} 
 \label{KL}
\begin{eqnarray}
\label{K}
 K[\lambda,\rho^{\star}]&=&   
 \int_0^{\infty}dx \; 4 \pi x^2 \; \overline{\Delta}_{\lambda}^2(x) \; 
 h_{\text{HS}, \; \rho}(x)
 \; , \\ 
 \label{L} 
  L[\lambda,\rho^{\star}]&=&
  \frac{1}{\rho^{\star}}	 \int_0^{\infty}dx \; 4 \pi x^2	
   \overline{\Delta}_{\lambda}^3(x)        \;,
\end{eqnarray}
\end{subequations}  
where $x=r/\sigma$ is the distance in reduced units and 
$ \overline{\Delta}_{\lambda}(x)\equiv
\beta^{\star}\rho^{\star}\sigma^{-2} \Delta_{\lambda}(r=\sigma x) $ is
the reduced propagator. 
We stress once again that, since  $\Delta_{\lambda}$ is simply related to 
the pair
potential $w$   (cf equations~(\ref{Deltak}), and~(\ref{Deltar})) 
it depends therefore on the regularization scheme. For a given regularization
scheme it depends
only upon the parameter $\lambda$.
Moreover, we  note that $\overline{\beta f}^{(2)}$ is
invariant under the transform $\xi \to 1/\xi$ as expected and that the second
term in the RHS of equation~(\ref{f2red}) vanishes in the case of the 
RPM ($\xi=1$). 
Expressions~(\ref{K}) 
and~(\ref{L}) can  easily be computed numerically for each regularization
scheme (for numerical details see Appendix~B).
%%%%%%%%%%%%%%%%%%%%%%%%%%%%%%%%%%%%%%%%%%%%%%%%%%%%%%%%
%%%%%%%%%%%%%%%%%%%%%%%%%%%%%%%%%%%%%%%%%%%%%%%%%%%%%%%%
\section{\label{LG}The liquid-vapor transition of the binary SPM}
\subsection{Theoretical background}
In past recent years extensive Monte Carlo simulations have been performed
to  study the liquid-vapor transition of the RPM and to elucidate the nature of
ionic criticality. It seems now well established that the critical exponents of
the RPM are those of the $3D$ Ising model \cite{Pana,CLW1,CLW2,CLW3}.
The more recent estimates of the Orsay group
for the  critical temperature, density, and chemical
potentials  are \cite{CLW3} $T^{\star}_c=0.049 17\pm 0.000 02$,
 $\rho^{\star}_c=0.080 \pm 0.005$, and $\nu^{\star}_c=-13.600\pm0.005$
 respectively (the critical pressure is not  known with accurate precision).
  Recent numerical studies are also available for 
 binary mixtures with an asymmetry in charge or/and
  in size \cite{Pana-Fisher,dePablo}.
 
 Here we consider only the binary SPM  and study various MF theories built 
 with the approximate free energies discussed in section~\ref{Binary}.
 We work in the framework of the Landau theory with a Landau function given by
 the expression \cite{Chaikin}
\begin{equation}
\label{w} 
w(\nu,\beta,\varphi)=\beta f(\beta,\varphi) -\nu \varphi \; ,
\end{equation}
the minimum of which gives minus the MF pressure, i.e.\cite{Chaikin}
\begin{equation}
\label{P-MF}
-\beta P_{\text{MF}}(\nu,\beta) = \min_{\varphi}w(\nu,\beta,\varphi) \; . 
\end{equation} 
Note that $\varphi$ plays the role of the order parameter of the Landau theory.
 The free energy function $\beta f(\beta,\varphi)$ 
which enters equation~(\ref{w}) can be rewritten as
\begin{equation}
\beta f(\beta,\varphi)=\beta f_{\text{HS}}(\varphi) -\varphi 
 \left( 
\frac{\xi}{1+\xi} \ln\xi  -\ln \left(1+\xi\right) 
  \right) + J(\lambda,\varphi) \; ,
\end{equation}
where 
\begin{equation}
\label{J}
J(\lambda,\varphi)=I(\lambda) - \frac{1}{4}K(\lambda,\varphi) +
\frac{1}{12} \frac{(1-\xi)^2}{\xi} L[\lambda,\varphi] \; .
\end{equation}
$\beta f_{\text{HS}}(\varphi)$ is the  free energy of the HS reference
system as given by equation~(\ref{fHS}) and
the functions $I$, $K$, 
and $L$ in the RHS of equation~(\ref{J}) are those defined
in section~\ref{Binary}.
Note that  we have dropped all the ''$\star$'' to simplify the notations, 
reduced
quantities are implicitly meant.

Above $T_c$ and for any chemical potential $\nu$,
  $w(\nu,\beta,\varphi)$ is a convex function of the order parameter 
$\varphi$ and equation~(\ref{P-MF}) admits a unique solution
 $\overline{\varphi}(\nu,\beta)$ which
is the (unique) solution of  the stationary condition
\begin{equation}
\label{statio} 
\left.\frac{\partial w}{\partial\varphi } \right|_{\overline{\varphi}}=0 \; .
\end{equation}
Therefore the MF pressure is given by $
\beta P_{\text{MF}}(\nu,\beta) = - w(\nu,\beta,\overline{\varphi}) 
$ which yields 
the expressions of the MF densities and energies per unit volume
 
 \begin{subequations} 
 \label{MF}
\begin{eqnarray}
\label{roMF}
 \rho_{\text{MF}}(\nu,\beta)&=& \frac{\partial\beta P_{\text{MF}} }
 {\partial \nu }    
 =\overline{\varphi} \; , \\
 \label{uMF} 
 u_{\text{MF}}(\nu,\beta)&=& - \frac{\partial\beta P_{MF} }
 {\partial \beta }=
 4 \pi \overline{\varphi} \; 
 \frac{\partial J }{\partial \lambda }(\lambda,\overline{\varphi}) 
         \;,
\end{eqnarray}
\end{subequations}  
where we made use of the stationary condition~(\ref{statio}).

\begin{figure}
\includegraphics[angle=270,scale=0.6]{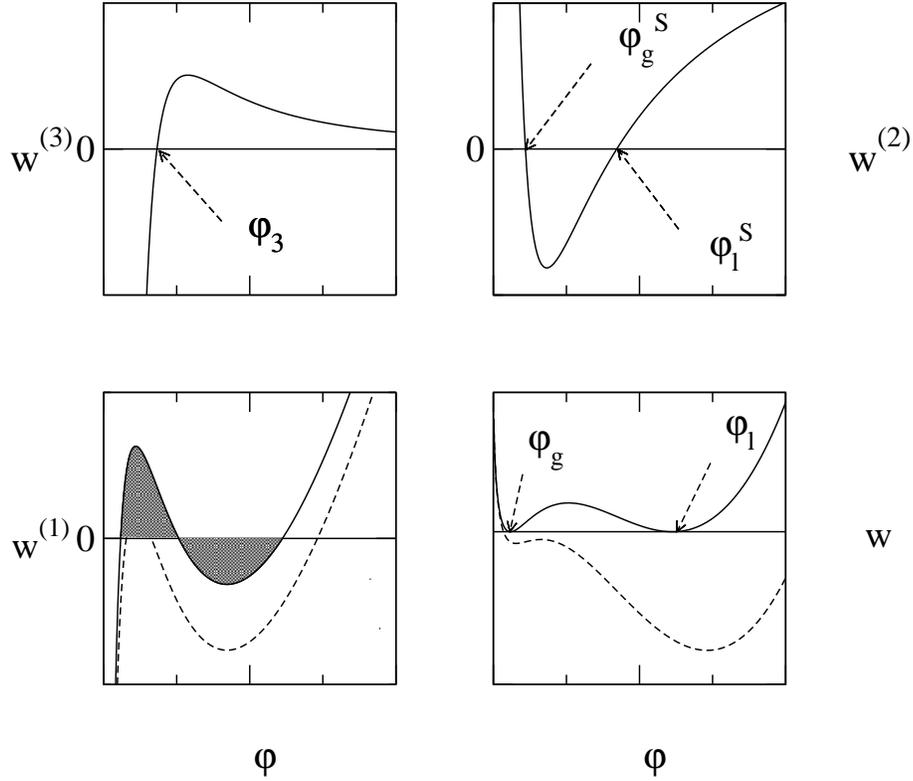}
\caption{\label{figw} The Landau function $w(\varphi)$ and its first
three derivatives $w^{(n)}(\varphi)$ at
$T<T_c$. The functions $w^{(2)}$ and $w^{(3)}$ are independent of the chemical
potential. The zeros of $w^{(2)}$ give the spinodal densities.  $w^{(1)}$
and  $w$ are diplayed for arbitrary $\nu$ (dashed curves) and for 
$\nu= \nu_{\text{coex}}$ (solid curves). At the coexistence the grayish areas
are equal.}
\end{figure}

Below $T_c$, $w(\nu,\beta,\varphi)$ is no longer a convex function of $\varphi$
and a second minimum of  $w(\nu,\beta,\varphi)$ arises for some range of 
chemical potentials. For $T<T_c$
($\beta > \beta_c$) there is a unique chemical potential
 $\nu_{\text{coex}}(\beta) $ 
corresponding to two values $\overline{\varphi}_g$ and $\overline{\varphi}_l
>\overline{\varphi}_g $
such that $-\beta P_{\text{MF}, \;\text{coex}}(\beta)=
w(\nu_{\text{coex}},\beta,\overline{\varphi}_g)=
w(\nu_{\text{coex}},\beta,\overline{\varphi}_l)$ (see figure~(\ref{figw})). 
These
two minima corresponds to the  coexistence of a gas phase at the density 
$\rho_{\text{MF}, \; g}(\beta)=\overline{\varphi}_g$ and a liquid at the 
density
$\rho_{\text{MF}, \; l}(\beta) =\overline{\varphi}_l$. 
At the critical temperature 
the two solutions $\overline{\varphi}_g$ and  $\overline{\varphi}_l$  
merge into a unique solution  $\overline{\varphi}_c=\rho_{\text{MF}, \; c}$ 
corresponding to the critical density.

We have determined numerically the coexistence curve of the SPM for 
the one-loop and two-loop expressions of the free energy for 
various regularizations schemes discussed in section~\ref{Prole}, i.e.
\begin{itemize}
\item charge smearing regularization scheme
(i.e.  $w(r) =w_{a}(r)$). We limited ourselves to the case
of the
surface distribution~(\ref{tau-a}) with $a^*=a/\sigma$ ranging from $0$ (DH case) to
$1$ (OMF case).
\item MSA regularization scheme (i.e.  $w(r) =w_{\text{MSA}}(r)$).
\item  WCA regularization scheme (i.e.  $w(r) =w_{\text{WCA}}(r)$,
cf sec.~(\ref{WCA})).
In this case a numerical solution was
obtained only for temperatures not to far below $T_c$ in order to avoid the 
RPA catastrophe.
\end{itemize}

\subsection{Numerical results}
We have found worthwhile to discuss shortly the  algorithm
devised to obtain the coexistence curve in appendix~B. The results for the
critical temperatures, chemical potentials, pressures, densities and energies 
are summarized
in tables~(\ref{one-loop},~\ref{two-loop},~\ref{assym}) and are discussed below.
%%%%%%%%%%%%%%%%%%%%%%%%%%%%%%%%%%%%%%%%%%%%%%%%%%%%%%%%%%%%%%%%%%%%%%%%%%%%%
%%%%%%%%%%%%%%%%%%%%%%%%%%%%%%%%%%%%%%%%%%%%%%%%%%%%%%%%%%%%%%%%%%%%%%%%%%%%%%
\subsubsection{One-loop results}

We have already noticed that the free energy of the binary SPM is independent
of the charge assymetry factor $\xi$ at the one-loop order. The coexistence
curves and notably  the critical
parameters ($T_c^*$, $\rho_c^*$, \ldots etc) are therefore independent
of  $\xi$ which  is a serious
drawback clearly in contradiction with numerical simulation results
\cite{Pana-Fisher,dePablo}.

\begin{figure}
\includegraphics[angle=270,scale=0.6]{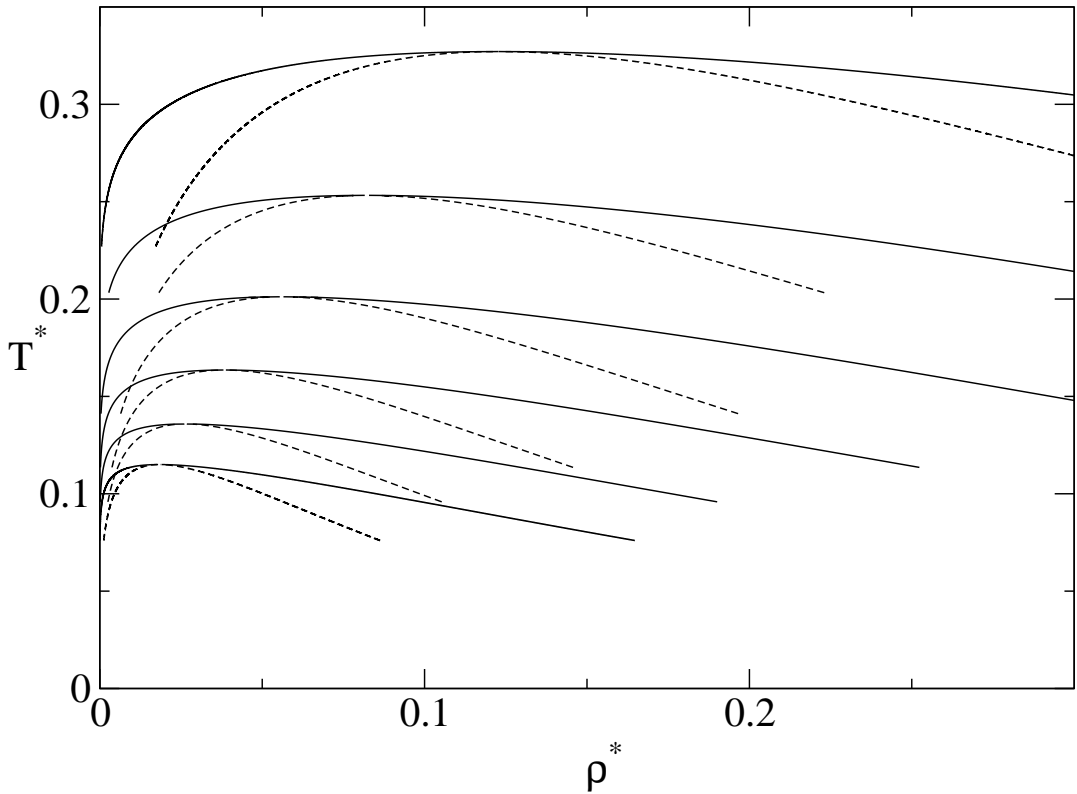}
\caption{\label{OMF} Coexistence curves of the RPM in the
one-loop approximation with  $w(r)=w_a(r)$. 
From top to bottom $a^*=0$ (DH), $a^*=0.2$, $a^*=0.4$, $a^*=0.6$, $a^*=0.8$,
 $a^*=1$ (OMF).
Solid lines : coexistence curves, dashed
line: spinodal curves.}
\end{figure}

Our data for the one-loop approximation are reported in table~(\ref{one-loop}).
In the case of the MSA regularization
scheme one recovers the MSA theory provided that 
the HS free energy is chosen to be that of the PY approximation. In this case
we observed a perfect agreement of our data those obtained by Gonzales-Tovar
\cite{Gonzales}. We stress that the results reported in 
table~(\ref{one-loop}) for the MSA  regularization scheme were obtained, as all
other results reported in all the tables of this paper, by
making use of the Carnahan-Starling approximation for the HS free energy and of
the PY solution $h_{\text{PY}}(r)$ of the HS corelation function.

\begin{figure}
\includegraphics[angle=270,scale=0.6]{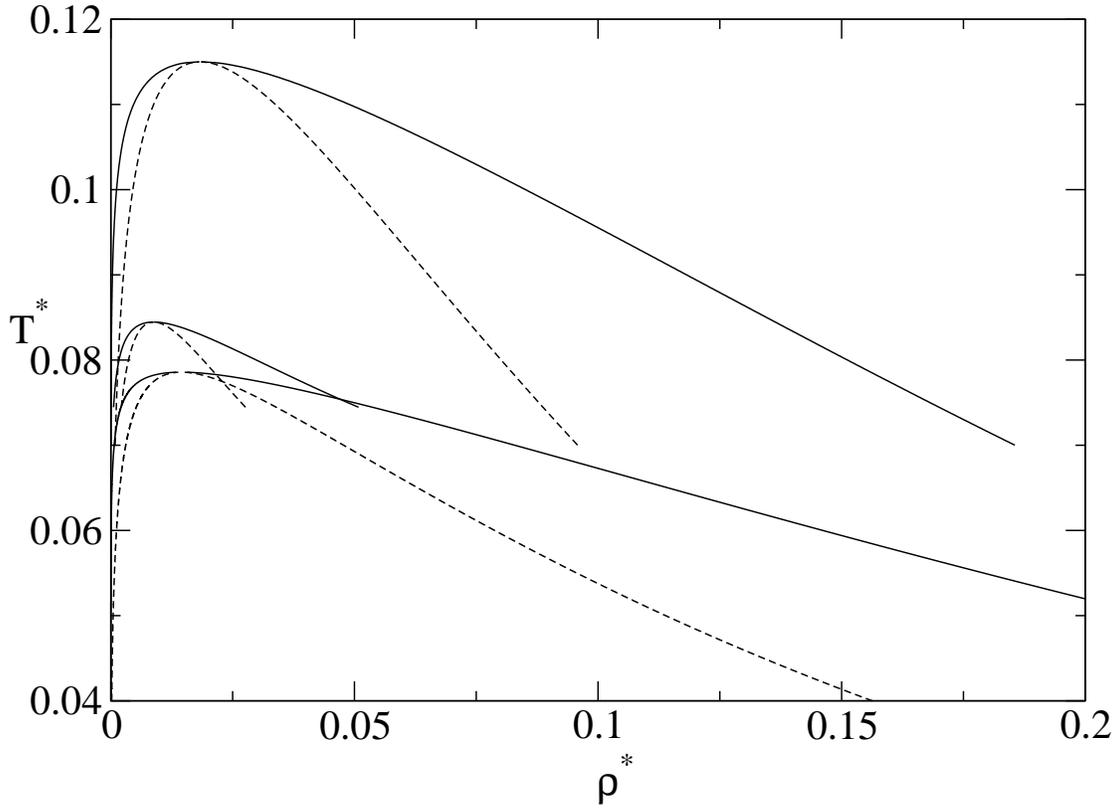}
\caption{\label{ONE} Coexistence curves of the RPM in the one-loop 
approximation. 
From top to bottom: OMF scheme, WCA
scheme, MSA scheme. 
Solid lines : coexistence curves, dashed
line: spinodal curves.}
\end{figure}

As apparent in table~(\ref{one-loop}) and figures~(\ref{OMF}) and~(\ref{ONE})
the coexistence curves depend strongly upon the regularization scheme adopted 
for the Coulomb potential inside the HS core. This is particularly striking for
the charge smearing regularization scheme where the critical temperature $T_c^*$
decreases steadily from $T_c^*=0.3271$  for $a^*=0$ (DH case) to $T_c^*=0.1150$ 
for $a^*\equiv a/\sigma=1$ (OMF case). 
Similarly the critical density $\rho_c^*$ decreases  with the diameter
$a$ of the charge distribution from  $\rho_c^*=0.12267$  for  $a^*=0$ to
$\rho_c^*=0.02198$  for  $a^*=1$.

The coexistence curves at the one-loop order obtained with the OMF, WCA,
and MSA (or ORPA) regularization schemes are displayed in figure~(\ref{ONE}). 
The MSA scheme gives the better critical temperature, i.e. 
$T_c^{* \; \text{MSA}}=0.07858$ to be compared with the exact $T_c^*=0.04917(2)$
\cite{CLW3},
whereas the critical density is largely underestimated, i.e.  
$\rho_c^{* \; \text{MSA}}=0.01449$ to be compared with the exact 
$\rho_c^{*}=0.080(5)$ \cite{CLW3}. By contrast the OMF scheme gives a slightly
better critical density than the MSA scheme - however still largely
underestimated-, i.e.  
$\rho_c^{* \; \text{OMF}}=0.01834$, but overestimates  $T_c^*$, i.e 
$T_c^{* \; \text{OMF}}=0.1150$. The results obtained in the framework of the
WCA scheme lie somewhere between those obtained via the OMF and MSA
schemes.

%%%%%%%%%%%%%%%%%%%%%%%%%%%%%%%%%%%%%%%%%%%%%%%%%%%%%%%%%%%%%%%%%%%%%%%%%%%%%
%%%%%%%%%%%%%%%%%%%%%%%%%%%%%%%%%%%%%%%%%%%%%%%%%%%%%%%%%%%%%%%%%%%%%%%%%%%%%%
\subsubsection{Two-loop results}

\begin{figure}
\includegraphics[angle=270,scale=0.6]{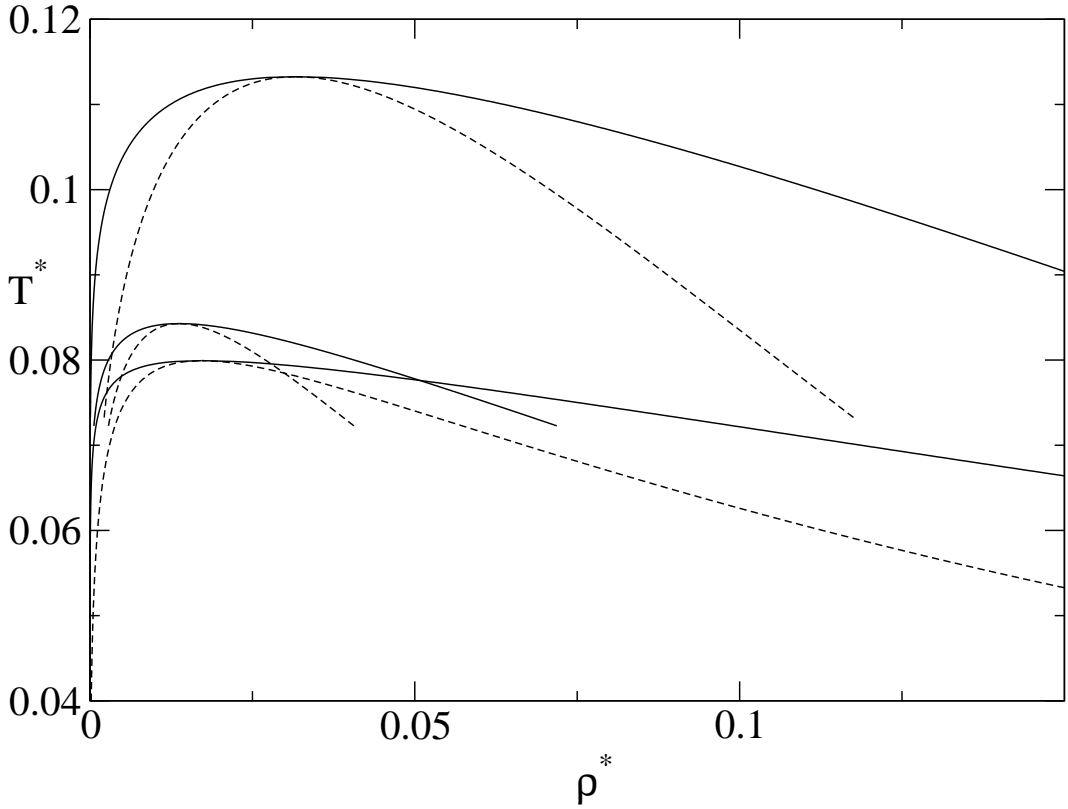}
\caption{\label{TWO}Coexistence curves of the RPM in
the two-loop approximation. 
From top to bottom: OMF scheme, WCA
scheme, MSA scheme. 
Solid lines : coexistence curves, dashed
line: spinodal curves.}
\end{figure}

The critical parameters of the RPM at the two-loop order are given in 
table~(\ref{two-loop}) and the coexistence curves are sketched in 
figure~(\ref{TWO}). Including the two-loop correction yields a slight
decrease of the
critical temperature in all cases (except for the MSA case) as well as an
increase of the critical density. More precisely one finds,
at the two-loop level, the following critical parameters :
$(T_c^{* \; \text{OMF}}=0.1133, \; \rho_c^{* \; \text{OMF}}=0.0316)$,
$(T_c^{* \; \text{MSA}}=0.07993, \; \rho_c^{* \; \text{MSA}}=0.01722)$,
and $(T_c^{* \; \text{WCA}}=0.08428, \; \rho_c^{* \; \text{WCA}}=0.0137)$ in the
OMF, MSA, and WCA schemes respectively.
The improvement upon one-loop results 
is therefore significant
but the theoretical predictions, notably those for the critical densities, are
still far from the MC results. Our theoretical results can be (unfavorably)
compared with those obtained recently in the framework of the the so-called
collective variable (CV) method developed by the Ukrainian school which yield
$(T_c^{*}=0.0502, \; \rho_c^{*}=0.042)$
\cite{Patsahan0,Patsahan2}. At the moment the relation between the CV
method and the KSSHE formalism is not established, 
work in this direction in under way.  

At the two-loop order the free energy of the binary SPM depends explicitely on
the charge assymetry factor $\xi$. The effects of $\xi$ on the coexistence
curve parameters in the OMF, WCA, and MSA schemes are resumed 
in table~(\ref{assym}) for $\xi=1,2,3,4,5$. In all the considered
regularization schemes the critical temperature $T_c^*$
is found to increase slightly with  $\xi$
in clear contradiction with MC simulation data where a rapid decrease 
of $T_c^*$ with $\xi$ has been observed \cite{Pana-Fisher}. However, as
apparent in table~(\ref{two-loop}), the critical densities -despite much too 
low values- are found to increase
slightly with $\xi$ in the OMF and MSA schemes, which is in qualitative
agreement with the MC data of ref.~(\onlinecite{Pana-Fisher}).

%%%%%%%%%%%%%%%%%%%%%%%%%%%%%%%%%%%%%%%%%%%%%%%%%%%%%%%%
%%%%%%%%%%%%%%%%%%%%%%%%%%%%%%%%%%%%%%%%%%%%%%%%%%%%%%%%
\section{\label{Conclusion}Summary}
In this paper we have studied the liquid-vapor coexistence curve of the SPM 
-i.e. a simple
version of the primitive model,  the
prototype of a system governed by long range Coulomb interactions-, by means of
various mean-field  theories. 
These theories were obtained in the framework of the KSSHE
field-theoretical representation of charged HS systems
by means of a loop expansion of the free energy of the homogeneous system
\cite{Cai-JSP}.
Some of these MF theories are
equivalent to well known approximations of the theory of liquids such that
the Debye-H\"uckel, or MSA (ORPA) theory, other were considered here for 
the first time. The results
obtained for the critical point are  in all cases in poor agreement with
the MC simulations. 

The peculiarity of ionic criticality is clearly seen in the
KSSHE formalism. The zero-loop order of the KSSHE theory
is unable to reproduce a liquid-vapor transition and provides merely 
the local charge
neutrality condition \cite{Cai-JSP}; the transition appears only
at the one-loop order which is an unusual feature and suggests some kind of
inconsistence of the KSSHE theory for ionic systems.
The same remark holds for the
other field-theoretical treatments of the
SPM \cite{Ciach1,Ciach2,Ciach3,Patsahan0,Patsahan2}. 
From the KSSHE side, further challenges are the derivation of the one-loop
corrections to the density correlation functions and generalization of the
two-loop expansions of the homogeneous free energy to the case of a PM 
with size assymetry.

%%%%%%%%%%%%%%%%%%%%%%%%%%%%%%%%%%%%%%%%%%%%%%%%%%%%%%%%%%%%
\begin{acknowledgments}
The author likes to thank O. Patsahan, I. Mryglod, A. Ciach, and B. Jancovici 
for useful discussions.
\end{acknowledgments}
%%%%%%%%%%%%%%%%%%%%%%%%%%%%%%%%%%%%%%%%%%%%%%%%%%%%%%%%%%%%
%%%%%%%%%%%%%%%%%%%%%%%%%%%%%%%%%%%%%%%%%%%%%%%%%%%%%%%%%%%%
\appendix
\section{\label{apA}The relations between 
$G^{(n)\; \text{T}}_\text{C}$ and $G^{(n)\; \text{T}}_{\varphi}$ }

We shall denote by $\Xi[ \{\nu_{\alpha} \},V_{\text{e}}]$ the GCPF
of the SPM in the presence of an external electrostatic potential
$V_{\text{e}}(1)$. The $M$ dimensionless chemical potentials $\nu_{\alpha}$
are assumed to be uniform.  
Let $\rho_{\text{e}}(1)$ be the
external charge distribution giving rise to $V_{\text{e}}(1)$. 
$V_{\text{e}}(1)$ and $\rho_{\text{e}}(1)$ are linked by the Poisson equation
which can be written, with the help of the formal notations
 of sec.~(\ref{Prole}), as
\begin{equation}
\label{Poisson} 
v_c^{-1}(1,2) V_{\text{e}}(2)=\rho_{\text{e}}(1) \; .
\end{equation}
 Obviously the 
 charge correlation functions defined at equation~(\ref{defcorreCT3})
 can  be expressed as the functional derivatives
\begin{eqnarray}
\label{GCdef}
G^{(n)}_\text{C}(1,\ldots,n)&=&
\frac{(-1)^n}{\beta^n}
 \frac{1}{ \Xi[ \{\nu_{\alpha} \}]} \; 
\left. 
\frac{\delta^n \Xi[ \{\nu_{\alpha} \},V_{\text{e}}] }
{\delta V_{\text{e}}(1) \ldots \delta V_{\text{e}}(n) }
\right \vert_{V_{\text{e}}=0} \; ,  \nonumber \\
G^{(n)\; \text{T}}_\text{C}(1,\ldots,n)&=&
\frac{(-1)^n}{\beta^n} \; 
\left. 
\frac{\delta^n \ln\Xi[ \{\nu_{\alpha} \},V_{\text{e}}] }
{\delta V_{\text{e}}(1) \ldots \delta V_{\text{e}}(n) }
\right \vert_{V_{\text{e}}=0} \; .
\end{eqnarray} 

Expressions similar to equations~(\ref{GCdef}) can be derived for the field correlation functions $G^{(n)}_{\varphi}$ and
$G^{(n)\; \text{T}}_{\varphi}$ in the following way. Firstly,
note that  one has $\Xi[ \{\nu_{\alpha} \},V_{\text{e}}]
\equiv \Xi[ \{\nu_{\alpha}-q_{\alpha} \overline{V}_{\text{e}} \}]$
 where the ''smeared'' field $\overline{V}_{\text{e}}$ is given by
 $\overline{V}_{\text{e}}(1)=\beta \tau(1,2) \; V_{\text{e}}(2)$.
Therefore
the expression~(\ref{Xinewlook}) of $\Xi$ can be written more explicitely
as 
\begin{equation}
\label{Xi2}
\Xi[ \{\nu_{\alpha}\},V_{\text{e}} ]=
 {\cal N}_{v_{c}}^{-1} \int {\cal D} \varphi
\; \exp(- \frac{1}{2} \left< 
\varphi \vert  v_{c}^{-1} \vert \varphi \right>  - 
\ln \Xi_{\text{HS}}\left[ \{\overline{\nu}_{\alpha} +i q_{\alpha}
(\phi -\overline{V}_{\text{e}}) \}\right])
\: .
\end{equation}
We now perform the change of variables 
\begin{equation}
\varphi \to \varphi -i\beta^{1/2} V_{\text{e}}
\end{equation}
in equation~(\ref{Xi2}). The functional Jacobian is equal to one and this 
gives us, after taking the logarithm 
\begin{equation}
\label{Xi3}
\ln \Xi[ \{\nu_{\alpha}\},V_{\text{e}} ]=
\frac{\beta}{2} \left< 
V_{\text{e}} \vert  v_{c}^{-1} \vert V_{\text{e}} \right> +
\ln \Xi^*[ \{\nu_{\alpha} \},B] \; ,
\end{equation}
where $\ln \Xi^*[ \{\nu_{\alpha} \},B]$ is the KSSHE partition 
function~(\ref{Xinewlook}) associated with the modified action
\begin{equation}
\label{XiB}
{\cal H}[\varphi] -B(1)\varphi(1) \; ,
\end{equation}
where $B(1)\equiv i \beta^{1/2}v_{c}^{-1}(1,2) V_{\text{e}}(2)$ plays the role
of an additional external magnetic field. As it is well known, the functionals 
 $ \Xi^*[ \{\nu_{\alpha}\},B ]$  and
$\ln \Xi^*[ \{\nu_{\alpha}\},B ]$ are the generators of the 
ordinary and connected correlation functions of the KSSHE field 
$\varphi$ respectively
\cite{Zinn}.
Therefore we have

\begin{eqnarray}
\label{Gphidef} 
G^{(n)}_{\varphi}(1,\ldots,n)&= & 
\frac{1}{\Xi[ \{\nu_{\alpha} \}] }
\left. 
\frac{\delta^n \Xi^*[ \{\nu_{\alpha} \},B] }
{\delta B(1) \ldots \delta B(n) }
\right \vert_{B=0}
\; \nonumber \\
G^{(n)\; \text{T}}_{\varphi}(1,\ldots,n)&= &
\left. 
\frac{\delta^n \ln\Xi^*[ \{\nu_{\alpha} \},B] }
{\delta B(1) \ldots \delta B(n) }
\right \vert_{B=0} \; .
\end{eqnarray}
Taking the functional derivative of both sides of equation~(\ref{Xi3}) with
 respect
to $V_{\text{e}}$ and noting that 
\begin{equation}
\label{zozo}
\frac{\delta B(1)}
{\delta V_{\text{e}}(2) }= i\beta^{1/2} v_{c}^{-1}(1,2) \; , 
\end{equation}
one finds that
\begin{eqnarray}
\label{aaa1}
\rho_\text{C}(1)+\rho_{\text{e}}(1)&=&-i \beta^{-1/2}v_{c}^{-1}(1,2)< \varphi(1)> 
\nonumber \\
&\equiv & \frac{i}{4 \pi \beta^{1/2}}  \Delta_{1} < \varphi(1)> 
\; ,
\end{eqnarray}
yielding the charge neutrality condition $\rho_\text{C}(1)=0$ for a uniform system 
in the
absence of external charge, since in that case $< \varphi(1)>=\text{ cte }$.

Taking twice the functional derivative of both sides
of equation~(\ref{Xi3})
with respect
to $V_{\text{e}}$ in the limit $B \to 0$  yields readily
\begin{equation}
\label{aaa2}
\beta G^{(2)\; \text{T}}_{\text{C}}(1, 2)= \frac{-1}{4 \pi} \Delta_{1} \delta(1,2)
 -  \frac{1}{(4 \pi)^{2}} \Delta_{1} \Delta_{2}
  G^{(2) \; \text{T}}_{\varphi}(1,2) \; ,
\end{equation}
where we  made use of equations~(\ref{GCdef}), (\ref{Gphidef}), and 
(\ref{zozo}).
Finally, continuing the process and
differentiating equation~(\ref{Xi3}) $n$ times ($n\geq 3$) one obtains
\begin{equation}
\label{aaa3}
\beta^{n/2} i^n G^{(n)\; \text{T}}_\text{C}(1,\ldots,n)= 
\frac{(-1)^{n}}{(4 \pi)^{n}} \Delta_{1} \ldots  \Delta_{n}
  G^{(n) \; \text{T}}_{\varphi}(1, \ldots,n)  \; .
\end{equation}
Equations~(\ref{aaa1}),~(\ref{aaa2}), and~(\ref{aaa3}) were obtained in a
complicated way in ref.~(\onlinecite{Cai-JSP}).
%%%%%%%%%%%%%%%%%%%%%%%%%%%%%%%%%%%%%%%%%%%%%%%%%%%%%%%%%%%%
%%%%%%%%%%%%%%%%%%%%%%%%%%%%%%%%%%%%%%%%%%%%%%%%%%%%%%%%%%%%
\section{\label{apB}An  algorithm to determine the coexistence curve}
In this appendix we give some details on the algorithm used to obtain the
coexistence curve of the SPM in the various approximation schemes
considered in the paper. 
We denote by $w^{(n)}(\nu,\beta,\varphi)$ the n-th partial
derivative of the Landau function with respect to the order parameter
$\varphi$. For $n\geq 2$,  $w^{(n)}(\nu,\beta,\varphi)$ is 
independent of the chemical potential $\nu$ as apparent from
 equation~(\ref{w}).
At any temperature $T$ the function  $w^{(3)}(\beta,\varphi)$ versus 
$\varphi$  has the shape depicted in
figure~(\ref{figw}), i.e. with a unique zero located at some $\varphi_3$.
A precise numerical estimate of $\varphi_3$ can be obtained
with the help of the bisection method.  In that aim  we used
the routine RTBIS of Numerical Recipes\cite{Numrep}. More precisely
\begin{equation}
\varphi_3=\text{RTBIS}(w^{(3)},\varphi_{\text{min}},
\varphi_{\text{max}},\epsilon) \; ,
\end{equation}
where RTBIS seeks for the (unique)  zero of $w^{(3)}(\varphi)$ in 
the prescribed interval
$[\varphi_{\text{min}},\varphi_{\text{max}}]$ with some arbitrary
precision $\epsilon$.

Therefore the function  $w^{(2)}(\beta,\varphi)$ versus 
$\varphi$ is convex for all $\beta$.
For $T<T_c$  $w^{(2)}$ has the shape depicted in
figure~(\ref{figw}), i.e. with a unique negative minimum at 
$\varphi=\varphi_3$. 
For $T>T_c$ the minimum is positive and thus the critical temperature $T_c$ 
is obtained by the condition $w^{(2)}(\beta_c,\varphi_3)=0$. 
The zeros of 
$w^{(2)}(\varphi)$ for $T<T_c$ will be denoted  $\varphi_g^S$ and 
$\varphi_l^S > \varphi_g^S$. They
are also easily obtained  with the help of the bisection method, i.e.
\begin{eqnarray}
\varphi_g^S &= &\text{RTBIS}(w^{(2)},\varphi_{\text{min}},\varphi_3,\epsilon)
  \; ,  \nonumber \\
\varphi_l^S &=& \text{RTBIS}(w^{(2)},\varphi_{3},
\varphi_{\text{max}},\epsilon)
\; .    \nonumber
\end{eqnarray}  
Since there the second derivative of the free energy with respect to $\varphi$
(i.e. the density) vanishes, the locus of the points
$(\varphi_g^S , \varphi_l^S)$ is the so-called  spinodal curve \cite{Chaikin}.

The points  $(\varphi_g^S , \varphi_l^S)$ 
correspond  to the maximum and the minimum of the function
 $w^{(1)}(\varphi)$ respectively. For an arbitrary chemical potential
 $\nu \neq
\nu_{\text{coex}}$ one is likely to find the dashed curves of
 $w^{(1)}(\varphi)$
and $w(\varphi)$ displayed in figure~(\ref{figw}). One wants to find the 
coexistence
chemical potential $\nu_{\text{coex}}$ such that the two minima of 
$w(\varphi)$ have the same value, i.e. to find for $w(\varphi)$ and
$w^{(1)}(\varphi)$ the solid curves  depicted in
figure~(\ref{figw}). This can be achieved conveniently and precisely 
by means of the following iterative process.

(i) First one needs an estimate of $\nu$ not too far from $\nu_{\text{coex}}$.
Denoting by  $w^{(1)}_{\text{min}}$ and $w^{(1)}_{\text{max}}$ the values of
 the
minimum and maximum of $w^{(1)}(\varphi)$ respectively
one adopts for $\nu$ the new
value $\nu \to \nu_0 = \nu + 
(w^{(1)}_{\text{\text{min}}}+w^{(1)}_{\text{max}})/2 $
which ensures
that now  $w^{(1)}_{\text{min}}=-w^{(1)}_{\text{max}}$. 

(ii) The  zeros of 
$w^{(1)}(\varphi)$ in the intervals $[\varphi_{\text{min}},\varphi_g^S]$
 and $[\varphi_l^S,\varphi_{\text{max}}]$  
 are rough estimates of the
coexistence densities. Denoting them by $\varphi_g^0$ and $\varphi_l^0$
respectively one has
\begin{eqnarray}
\varphi_g^0 &= &\text{RTBIS}(w^{(1)},\varphi_{\text{min}},\varphi_g^S,\epsilon)
  \; ,  \nonumber \\
\varphi_l^0 &=& \text{RTBIS}(w^{(1)},\varphi_l^S,\varphi_{\text{max}},\epsilon)
\; .    \nonumber
\end{eqnarray}

(iii) At this point we have found a $\nu_0$ and two estimates 
$\varphi_g^0$ and $\varphi_l^0$ such that $w^{(1)}(\nu_0,\varphi_g^0)=
w^{(1)}(\nu_0,\varphi_l^0)=0$ but  $w(\nu_0,\varphi_g^0) \neq
w(\nu_0,\varphi_l^0)$ in general. It follows from equation~(\ref{w}) that if
one defines
\begin{equation}
\nu_1=\nu_0 + \frac{w(\nu_0,\varphi_l^0)- w(\nu_0,\varphi_g^0)}
{\varphi_l^0-\varphi_g^0 } \; ,
\end{equation} 
then $w(\nu_1,\varphi_g^0)=w(\nu_1,\varphi_l^0)$ but  
$w^{(1)}(\nu_1,\varphi_{g, l}^0)\neq 0$. One then 
determines 
two new estimates of $\varphi_l$ and $\varphi_g$ as
\begin{eqnarray}
\varphi_g^1 &= &\text{RTBIS}(w^{(1)}(\nu_1,\varphi),\varphi_{\text{min}},
\varphi_g^S,\epsilon)
  \; ,  \nonumber \\
\varphi_l^1 &=& \text{RTBIS}(w^{(1)}(\nu_1,\varphi),\varphi_l^S,
\varphi_{\text{max}},\epsilon)
\; .    \nonumber
\end{eqnarray}
 Now one has $w^{(1)}(\nu_1,\varphi_g^1)= w^{(1)}(\nu_1,\varphi_l^1)=0$ and one
 computes again $w(\nu_1,\varphi_g^1)$ and $ w(\nu_1,\varphi_l^1)$. If 
 \begin{equation}
 \vert \frac{w(\nu_1,\varphi_g^1)- w(\nu_1,\varphi_l^1)}
 {w(\nu_1,\varphi_g^1+ w(\nu_1,\varphi_l^1)} \vert < \epsilon
 \; ,
 \end{equation}
where $  \epsilon$ is the wanted precision, then the problem is numerically
solved. If it is not the case then one sets $\nu_1 \to \nu_0 $,  
$\varphi_{g,l}^1 \to \varphi_{g,l}^0$
 and one goes
back to point (iii). For any reasonable $\epsilon$ a few iterations are
necessary to get the result with a precision $  \epsilon$.

We end this appendix
by a few comments on the numerical calculation of $w$ 
and its derivatives $w^{(n)}$.
The HS free energy and its derivatives were computed in the Carnahan-Starling
approximation \cite{Hansen}. We also tried the PY expressions (in the
compressibility route) which makes little difference at the rather low 
densities
considered here.
Only the numerical estimate of the functions $K[\lambda,\varphi]$ 
(cf equation~(\ref{K})) 
which enter the free
energy at the second-loop order needs additional
comments.
In order to compute
$K[\lambda,\varphi]$ 
and its partial derivatives with respect to $\varphi$ (note that 
$\lambda= 4 \pi \beta \varphi$ does
also depends on $\varphi$)
we have used the PY solution for 
$h_{\text{HS}, \;\varphi}$. For numerical purposes it is necessary
to reexpress $K$ in terms of 
the function $\gamma_{\text{HS}, \;\varphi}(r)=
h_{\text{HS}, \;\varphi}(r)-c_{\text{HS}, \;\varphi}(r)$ 
which is a
continuous function of $r$. In the framework of PY theory analytical 
expressions can be
obtained for the Fourier transform $\widetilde{\gamma}_{\text{HS}, \;\varphi}$
 and its
derivatives with respect to $\varphi$
with the help of a  computer algebra package.
 $\gamma_{\text{HS}, \;\varphi}(r)$ and its derivatives  with respect
  to $\varphi$ are
then obtained as inverse numerical Fourier transforms.

%%%%%%%%%%%%%%%%%%%%%%%%%%%%%%%%%%%

\newpage
%%%%%%%%%%%%%%%%%%%%%%%%%%%%%%%%%%%%%%%%%%%%%%%%%%%%%%%%%%%%%%%%%%
\begin{table}[t!]
\caption{\label{one-loop} One-loop results for the critical temperature $T_c$,
chemical potential $\nu_{c}$, pressure $P_c$, density $\rho_c$, 
and energy $u_c$
of the SPM.
Various regularization schemes are considered, as explained in the text.}
\begin{ruledtabular}
\begin{tabular}{|r|d|d|d|d|d|} 
                 &  T_c         &  \nu_{c}     & 10^{2} \times   \beta P_c &
  \rho_c &   10 \times  u_c   
		  \\ \hline		  		   
\text{DH}&
  0.3271 &  -5.5274  & 2.4194  &    0.12267           &  -1.3316
\\   \hline
$ \tau_a \; :  \; a^*=0.1 $&
  0.2868 &  -5.8553  & 1.8336  &    0.10015           &  -0.9552
 \\   \hline 
$ a^*=0.2$&
  0.2532 &  -6.1688  & 1.4052  & 0.08224    &  -0.6981 
 \\   \hline
$ a^*=0.3$&
  0.2250 &  -6.4703 & 1.0860   & 0.06778    &  -0.5170 
 \\   \hline
  $ a^*=0.4$&
  0.2012 &  -6.7608 &   0.8453 &  0.05598  &  -0.3866
 \\   \hline 
   $ a^*=0.5$&
  0.1809 & -7.0408  &  0.6622  &	 0.04630  &  -0.2913
 \\   \hline 
   $ a^*=0.6$&
  0.1636   &  -7.3104 & 0.5220 & 0.03835	&  -0.2210
 \\   \hline
   $ a^*=0.7$&
 0.1487    &  -7.5695 & 0.4141 & 0.03180	&  -0.1686
 \\   \hline
   $ a^*=0.8$&
   0.1358  & -7.8182  & 0.3307  & 0.02641 &  -0.1293
 \\   \hline
    $ a^*=0.9$&
   0.1247  &  -8.0564  &  0.2659  & 0.02198  &   -0.09977
 \\   \hline
\text{OMF} &
  0.1150    &  -8.2843  &  0.2154 & 0.01834  &  -0.07745
\\   \hline 
\text{WCA}&
  0.08446    &  -9.1737  & 0.0930 & 0.00880 &  -0.02813
 \\   \hline 
\text{MSA}&
 0.07858     &  -9.1393 & 0.1229 & 0.01449 &  -0.04864
 \\       
\end{tabular}
\end{ruledtabular}
\end{table}
\newpage
%%%%%%%%%%%%%%%%%%%%%%%%%%%%%%%%%%%%%%%%%%%%%%%%%%%%%%%%%
\begin{table}[b!]
\caption{\label{two-loop}  Two-loop results for the critical 
temperature $T_c$,
chemical potential $\nu_{c}$, pressure $P_c$, density $\rho_c$,
and energy $u_c$
of the RPM.
Various regularization schemes are considered, as explained in the text.}
\begin{ruledtabular}
\begin{tabular}{|r|d|d|d|d|d|} 
                 &  T_c         &  \nu_{c}     & 10^{2} \times  \beta P_c &
  \rho_c &   10 \times  u_c   
		  \\ \hline		  	  		   		  		  		  	  		   
 \text{OMF} &
  0.1133 &  -7.8229  & 0.4181  &    0.0316           & -0.1363
\\   \hline
\text{WCA}&
  0.08428 &  -8.9142  & 0.1431  &    0.0137          & -0.0461
\\   \hline
\text{MSA}&
 0.07993     &  -9.0623 & 0.1379 & 0.01722 &  -0.06021
 \\       
\end{tabular}
\end{ruledtabular}
\end{table}
%%%%%%%%%%%%%%%%%%%%%%%%%%%%%%%%%%%%%%%%%%%%%%%%%%%%%%%%%
\newpage
\begin{table}[t!]
\caption{\label{assym} Two-loop results for the critical temperature $T_c$,
chemical potential $\nu_{c}$, pressure $P_c$, density $\rho_c$, and energy $u_c$
of the assymetric SPM.
Various regularization schemes are considered, as explained in the text.}
\begin{ruledtabular}
\begin{tabular}{|r|d|d|d|d|d|}
&  T_c         &  \nu_{c}     & 10^{2} \times   \beta P_c&
  \rho_c &   10 \times  u_c   
		  \\ \hline 
		  $\text{OMF} $ $\; \xi = 1$ &
  0.1133 &  -7.8229  & 0.4181  &    0.0316           & -0.1363
\\  \hline
$\text{OMF} $ $\; \xi = 2$ &
  0.1263 &  -7.0281  & 0.5848  &    0.0336           & -0.1209
\\ \hline 
$\text{OMF} $ $\; \xi = 3$ &
  0.1433 &  -6.1618  & 0.8049  &    0.0367           & -0.0966
\\ \hline
$\text{OMF} $ $\; \xi = 4$ &
  0.1585 &  -5.5175  & 1.0011  &    0.0396           & -0.0691
\\ \hline
$\text{OMF} $ $\; \xi = 5$ &
  0.1719 &  -5.0252  & 1.1744  &    0.0421           & -0.0399
\\ \hline
\text{WCA} $\; \xi = 1$ &
 0.08428 &  -8.9142  & 0.1431  &    0.0137          & -0.0461  
\\  \hline
\text{WCA} $\; \xi = 2$ &
 0.09334 &  -8.2044  & 0.1903  &    0.0132           & -0.0375
\\ \hline 
\text{WCA} $\; \xi = 3$ &
  0.1058 &  -7.4035  & 0.2582  &    0.0137           & -0.0305
\\ \hline
\text{WCA} $\; \xi = 4$ &
  0.1171 &  -6.8020  & 0.3208  &    0.0146           & -0.0244
\\ \hline
\text{WCA} $\; \xi = 5$ &
  0.1272 &  -6.3422  & 0.3770  &    0.0155           & -0.0183
\\ \hline
\text{MSA} $\; \xi = 1$ &
 0.07993     &  -9.0623 & 0.1379 & 0.0172 &  -0.0602  
\\  \hline
\text{MSA} $\; \xi = 2$ &
  0.08422 &  - 8.4828 & 0.1926  &    0.0199           & -0.0628
\\ \hline 
\text{MSA} $\; \xi = 3$ &
  0.09060 &  -7.7191 & 0.2751  &    0.0231           & -0.0602
\\ \hline
\text{MSA} $\; \xi = 4$ &
  0.09677 &  -7.0701  & 0.3555  &    0.0257           & -0.0527
\\ \hline
\text{MSA} $\; \xi = 5$ &
  0.1025&  - 6.5333 & 0.4302  &    0.0279           & -0.0422
\\ 

\end{tabular}
\end{ruledtabular}
\end{table}
%%%%%%%%%%%%%%%%%%%%%%%%%%%%%%%%%%%%%%%%%%%%%%%%%%%%%%%%%%%%%%%%%%%%%%%%%%%%%%

\end{document}